\documentclass[aps,prmaterials,superscriptaddress,preprint,onecolumn]{revtex4-2}

\usepackage{multirow}
\usepackage{graphicx}
\usepackage{amsfonts,amsmath,amssymb,amsthm}
\usepackage{epstopdf}
\usepackage{upgreek,xspace}
\usepackage{chngcntr}
\usepackage{xr}
\usepackage{ulem}
\usepackage[colorlinks,allcolors=blue]{hyperref}
\usepackage[version=3]{mhchem} 

\newcommand{\fig}[1]{Fig.~\ref{Fig#1}}

\begin{document}
\bibliographystyle{apsrev4-2}

\title{Relativistic Effects in \ce{LaBi2} Thin Films} 

\author{Reiley~Dorrian}
\affiliation{Department of Applied Physics and Materials Science, California Institute of Technology, Pasadena, California 91125, USA.}
\affiliation{Institute for Quantum Information and Matter, California Institute of Technology, Pasadena, California 91125, USA.}

\author{Sungmin~Song}
\affiliation{Department of Physics, Pohang University of Science and Technology, Pohang 37673, Republic of Korea}
\affiliation{Department of Physics and W. M. Keck Computational Materials Theory Center, California State University Northridge, Northridge,
California 91330, USA.}

\author{Jinwoong~Kim}
\affiliation{Department of Physics and W. M. Keck Computational Materials Theory Center, California State University Northridge, Northridge,
California 91330, USA.}
\affiliation{Korea Institute for Advanced Study, Seoul 02455, Republic of Korea}

\author{Mizuki~Ohno}
\affiliation{Department of Applied Physics and Materials Science, California Institute of Technology, Pasadena, California 91125, USA.}
\affiliation{Institute for Quantum Information and Matter, California Institute of Technology, Pasadena, California 91125, USA.}

\author{Seung-Hoon Jhi}
\affiliation{Department of Physics, Pohang University of Science and Technology, Pohang 37673, Republic of Korea}

\author{Nicholas~Kioussis}
\affiliation{Department of Physics and W. M. Keck Computational Materials Theory Center, California State University Northridge, Northridge,
California 91330, USA.}

\author{Joseph~Falson}
\email{falson@caltech.edu}
\affiliation{Department of Applied Physics and Materials Science, California Institute of Technology, Pasadena, California 91125, USA.}
\affiliation{Institute for Quantum Information and Matter, California Institute of Technology, Pasadena, California 91125, USA.}

\begin{abstract}
Chemical substitution in crystalline quantum materials is a powerful way to explore the consequences of strong spin-orbit coupling on their structural and electronic properties. In this work, we present an investigation of thin films of the \ce{La\textit{Pn}2} (\textit{Pn}~=~Sb, Bi) class of layered square-net intermetallics. We report the growth of \ce{LaBi2} with a pristine layer-by-layer growth mode, classifying it as a good metal displaying superconductivity at $\sim$0.55~K. Compared to \ce{LaSb2}, we attribute the enhanced metallic behavior and improved growth dynamics of \ce{LaBi2} to significant relativistic corrections to its electronic band structure and the resulting impact on both surface energy and intrinsic phonon scattering. 
\end{abstract}

\maketitle
\section{Introduction}

Spin-orbit coupling (SOC) -- the relativistic coupling of an electron's momentum to its spin -- is a mechanism of great importance for discovering and engineering novel electronic phases in crystalline materials. Under certain symmetry constraints, SOC acts to gap out spin-degenerate crossings in the electronic band structure, leading to inverted bands and symmetry-protected topological states \cite{hasan:review,qi:review}. In the superconducting state, strong SOC can impart significant resilience to applied magnetic fields through both scattering processes \cite{klb:1975} and momentum-dependent spin stiffness in the band structure, also known as Ising pairing \cite{zhou:2016}, while potentially enabling unconventional pairing symmetries through the mixing of spin-singlet and -triplet components \cite{youn:2012}. 

\begin{figure}[ht]
    \centering
    \includegraphics[width=85mm]{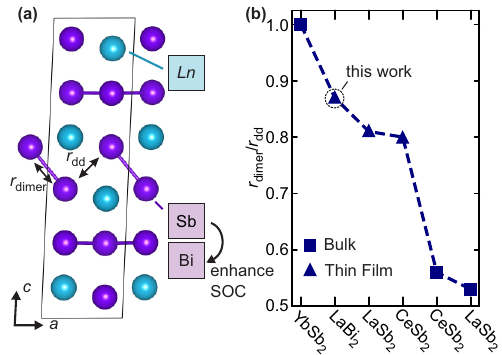}
    \caption{(a) Yb-monoclinic structure observed in thin films of the light rare-earth dipnictides. Substitution of the Sb anion with Bi is expected to enhance SOC due to the larger atomic mass. (b) Degree of spacer-layer \textit{Pn} dimerization in several reported compounds \cite{wang:1967}, including the results presented in this work.}\
    \label{Fig1}
\end{figure}

As the strength of SOC is related to the atomic mass of constituent atoms, chemical substitution in crystalline compounds is a commonly employed method for engineering the impacts of SOC \cite{hasan:review,qi:review,barrera:2018}. Exploring this degree of freedom has been fruitful in compounds containing the pnictogens Sb or Bi, where substitution of the former with the latter is found to fundamentally influence the topology of the electronic ground state \cite{liu:2013,liu:2016,he:2015,glazkova:2022}. However, investigations of the more subtle ways in which SOC influences a material's basic characteristics -- its growth kinetics, crystal structure, or metallic transport properties -- have received less attention.

In pursuit of this perspective, this work targets a class of compounds known for a chemistry-sensitive competition between structural and electronic phases -- the \ce{\textit{LnPn}2} (\textit{Ln} = lanthanide element, \textit{Pn}  = Sb, Bi) family of layered square-net intermetallics. A schematic representation of this structural class is shown in \fig{1}a. Variations in the oxidation state of the \textit{Ln} cation influence the stacking configuration of the corrugated \ce{\textit{Ln-Pn}} and quasi-planar \ce{\textit{Pn}} layers, resulting in distinct orthorhombic or monoclinic structure types which can be parametrized via the dimerization of \ce{\textit{Pn}} atoms within the corrugated layer (\fig{1}b)  \cite{papoian:2000}. In addition to the chemistry of the \textit{Ln} site, previous work on thin films of \ce{LaSb2} and \ce{CeSb2} discovered an instability with respect to Sb deficiency, establishing UHV thin-film growth techniques as a way to access new polymorphs with modified electronic properties \cite{llanos:2024,dorrian:2025}. 

With a desire to study the impacts of anion-tunable SOC in this subset of materials, this report builds upon the aforementioned work on (La, Ce)\ce{Sb2} \cite{llanos:2024,dorrian:2025} and focuses on thin-films of \ce{LaBi2} synthesized via molecular beam epitaxy (MBE). The substitution of \ce{Sb} with \ce{Bi} in this work acts to amplify SOC and significantly influences the thin-film growth mode and conductivity. We find \ce{LaBi2} to exhibit superior metallic behavior compared to \ce{LaSb2}, although not due to an intuition based on augmented hopping amplitudes, but rather due to suppressed phonon scattering at elevated temperatures arising from the spin-orbit induced shifting of electronic states. 

\section{Methods}
\subsection{Experimental Details}
We have synthesized \ce{LaBi2} thin films using an MBE system with a base pressure of $10^{-10}$ mbar on MgO (001) substrates which are laser-annealed as in Refs.~\cite{llanos_supercell:2024,llanos:2024,dorrian:2025} to optimize surface quality before growth. La and Bi fluxes were provided from conventional effusion cells with a large Bi/La beam-flux ratio (10$\sim$15) while the substrate was heated with a standard SiC heating coil. A growth rate-controlling La flux of about $7.5\times10^{-9}$ mbar established a growth rate of approximately 0.034 \AA/s. Due to the known air-sensitivity of the compound, all films were finalized with an amorphous Ge capping layer of approximately 5~nm thickness deposited \textit{in-situ} at room temperature and then stored in a \ce{N2}-atmosphere glovebox. Even with these precautions, film degradation was observed within a few hours of air exposure (See Fig.~S1 of the Supplemental Material~\cite{Supplement}).

Structural properties of the films were characterized by X-ray diffraction (XRD) (SmartLab, Rigaku). Electrical data above 2 K were collected using a Quantum Design Dynacool PPMS with a $14$ T superconducting magnet. The film was cut into an approximately square-shaped piece ($\approx 1$ - $2$~mm side lengths) and wire-bonded into a four-point van der Pauw geometry with Al wire. Magnetoresistance and Hall resistivity were extracted by symmetrizing the computed $\rho_\mathrm{xx}$ or anti-symmetrizing the diagonal $R_{yx}$ Van der Pauw channel, respectively, with respect to a magnetic field applied perpendicular to the film. Measurements of the superconducting transition below 1~K were gathered using a Leiden dilution refrigerator equipped with a two-axis vector magnet (9 and 3~T). The sample is fashioned into a rectangular ($\sim$1$\times$2~mm) piece and bonded with a linear four-point bond geometry.

\subsection{Theoretical Details}
Density functional theory (DFT) calculations were performed using the Quantum ESPRESSO package~\cite{giannozzi_quantum_2009}. The exchange-correlation interaction was described by the PBEsol functional~\cite{perdew_restoring_2008}, chosen for its accuracy in reproducing experimental lattice constants.
Ion-electron interactions were treated using norm-conserving pseudopotentials~\cite{hamann_optimized_2013}.
The electronic wavefunctions were expanded in a plane-wave basis set with a kinetic energy cutoff of 100~Ry.
To account for the partial occupancies of electronic states near the chemical potential, we employed the Methfessel-Paxton smearing technique~\cite{methfessel_high-precision_1989} with a broadening width of 0.01~Ry.
The Brillouin zone integration was performed using the Monkhorst-Pack scheme~\cite{monkhorst_special_1976}, where we used a $\mathbf{k}$-grid of $8 \times 6 \times 6$ for the bulk calculations. 
For the slab models used in the surface energy calculations, an $8 \times 8 \times 1$  $\mathbf{k}$-grid was adopted, and a vacuum layer of approximately $15\,\text{\AA}$ was inserted to avoid spurious interactions between periodic images. 
All atomic structures were fully relaxed until the residual forces on each atom converged to less than $1.0 \times 10^{-6}\,\text{Ry}/a_0$.

The electronic structure and electron-phonon coupling calculations were performed using interface with Wannier90~\cite{mostofi_wannier90_2008} and EPW code~\cite{ponce_epw_2016}. The elements of the electron-phonon matrix were initially computed on coarse grids of $8 \times 6 \times 6$ $\mathbf{k}$-points and $4 \times 2 \times 2$ $\mathbf{q}$-points, which were then interpolated onto dense $\mathbf{k}$- and $\mathbf{q}$-grid of $30 \times 20 \times 20$ for electron and phonon wavevectors, respectively. The electron group velocities were calculated using Wannier90, while the conductivity and scattering rates were obtained using EPW based on the Boltzmann Transport Equation (BTE). Integration for conductivity and scattering rates was performed using adaptive smearing within an energy window of $\pm 0.5$ eV around the Fermi level, ensuring sufficient coverage of the active transport states.

\begin{figure}[ht]
    \centering
    \includegraphics[width=85mm]{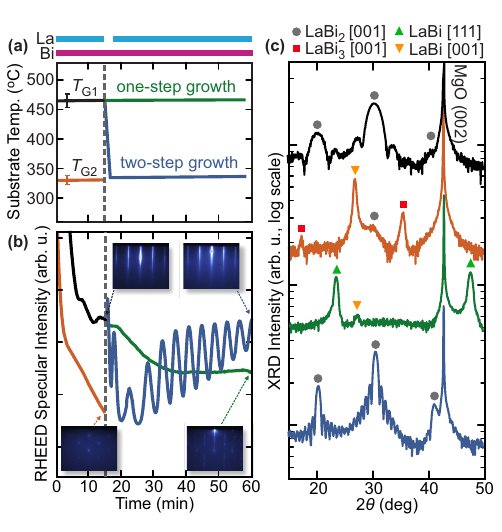}
    \caption{(a) Schematic view of growth approaches attempted for \ce{LaBi2} synthesis, showing substrate temperature over time with the provided elemental fluxes represented by colored bars at the top. (b) Intensity of specular RHEED diffraction over the course of growth for each approach. Inset: RHEED patterns at selected stages of growth. (c) Snippet of the X-ray diffraction (XRD) $\theta/2\theta$ diffraction pattern resulting from each growth approach.}
    \label{Fig2}
\end{figure}

\begin{figure}[ht]
    \centering
    \includegraphics[width=85mm]{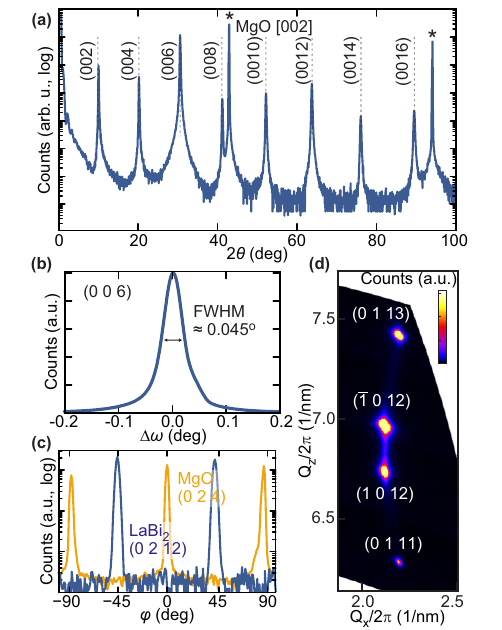}
    \caption{(a) $\theta/2\theta$ diffraction pattern from a 75 nm film produced by the optimized two-step growth approach. (b) Rocking curve of the (006) peak. (c) Azimuthal $\varphi$-scan of an asymmetric diffraction peak showing a [100]$_{\text{LaBi}_2}\parallel$ [110]$_{\text{MgO}}$ register with the MgO substrate.  (d) Reciprocal space map along the $h=1$ rod, showing finite splitting along $Q_z$ consistent with the Yb-mono structure.}
    \label{Fig3}
\end{figure}

\section{Properties of $\textbf{\ce{LaBi2}}$ Thin Films}
We begin by canvassing the history of \ce{LaBi2}. Previous reports have provided conflicting characterizations of the crystal structure, ranging from trigonal to orthorhombic crystal classes, and the structure is absent in the Inorganic Crystal Structure Database (ICSD)~\cite{hulliger:1977,nomura:1977,petrovic:2002}. This is presumably due in part to the extreme air sensitivity of the compound. Nevertheless, a layered compound reminiscent of \ce{LaSb2} is typically assumed to describe the structure. Despite being an excellent metal, \ce{LaBi2} remains the only member of the La-Bi alloy system without reported ambient-pressure superconductivity besides the rock-salt structured LaBi~\cite{hulliger:1977}. Meanwhile, in addition to the modified dimerization illustrated in \fig{1}b, thin films of both \ce{LaSb2}~\cite{llanos:2024} and \ce{CeSb2}~\cite{dorrian:2025} display a unique structure-type characterized by a monoclinic shear, in contrast to bulk orthorhombic crystals. The unique insights that thin films have provided into the delicate structural properties of this class of materials have motivated our efforts to synthesize \ce{LaBi2} and characterize its structure.

Figure \ref{Fig2} illustrates our entry point for producing high-quality, single-phase films. Four representative films correspond to distinct sequences of substrate temperatures (labeled $T_\mathrm{G1}\approx 450-465^\circ$C and $T_\mathrm{G2}\approx 325-335^\circ$C, \fig{2}a) with fixed La and Bi beam fluxes. We focus first on the initial 15 minutes of growth. At the higher temperature $T_\mathrm{G1}$ (black), the film quickly develops a streaky RHEED pattern with oscillating intensity (\fig{2}b), indicating a flat crystal coating the MgO surface in a layer-by-layer fashion. Out-of-plane XRD shows reflections from the (001) \ce{LaBi2} lattice planes without impurity phases, where we have assumed the same diffraction conditions as observed in bulk crystals (\fig{2}c)~\cite{yoshihara:1975,zhou:2018}. This higher 
$T_\mathrm{G}$, however, is found to be unsuitable for prolonged growth; the RHEED pattern gradually dims and degrades into a spottier series of streaks, as shown in \fig{2}b. XRD confirms this to represent a complete decomposition of the \ce{LaBi2} film into (111)-oriented LaBi, as shown in the green trace in \fig{2}c. This can be understood in terms of the continuous thermal desorption of Bi from the substrate; due to its low melting point and high vapor pressure, lower growth temperatures are typically employed in the initial stages of growth for thin films of Bi-based compounds to ensure proper Bi adhesion, followed by a higher $T_\mathrm{G}$ deposition to improve film crystallinity~\cite{li:2010,fan:2020,bansal:2011,ueda:2020,loke:2023}.

To test this approach, we begin with a lower temperature $T_\mathrm{G2}$ as illustrated by the orange traces in Figs.~\ref{Fig2}a-\ref{Fig2}c. Here, RHEED dims rapidly and displays diffuse features even at the 15 minute mark, with XRD reflecting a mixed-phase film composed of \ce{LaBi}[002], \ce{LaBi2}, and peaks associated with \ce{AuCu3}-structured \ce{LaBi3}~\cite{kinjo:2016}. Some combination of limited surface migration energy and excessive Bi incorporation at low temperatures likely contributes to this behavior, driving the system towards multiple competing phases. The approach to realizing single-phase films is therefore a two-step process; starting with a high-$T_\mathrm{G}$ buffer to obtain a phase pure template layer followed by an extended low-$T_\mathrm{G}$ growth period (blue), the RHEED pattern persists and oscillates in intensity, and the resulting film is found to be phase-pure \ce{LaBi2} with sharp (001) Bragg peaks and pronounced Laue fringes. This modified two-step procedure demonstrates the complex thermodynamics at play in the thin-film growth of intermetallic phases containing volatile species.

\begin{table*}[ht]
	\caption{\label{tab:bulk}
    Calculated relative total energies, lattice constants, and angles of the Sm-type and Yb-mono bulk phases of \ce{LaBi2} without (with) SOC. SG denotes the space group. 
    Experimental values reported in the literature and obtained in this work are also included for comparison.
    }
	\begin{ruledtabular}
		\begin{tabular}{l r|cr|lcccccc}
			& System & SG & $\Delta E$ (meV/La) & $a$ (\AA) & $b$ (\AA) & $c$ (\AA) & $\alpha$ ($^\circ$) & $\beta$ ($^\circ$) & $\gamma$ ($^\circ$) \\
			\hline
            
			\multirow{2}{*}{Without SOC} & Sm-type & 64 & 110.21 & 4.575 & 4.575  & 18.275 & 90.00 & 90.00 & 90.74 \\
			 & Yb-mono & 12 & 0.000 & 4.649 & 4.588 & 17.457 & 90.00 & 87.11 & 90.00 \\
            \hline
            \multirow{2}{*}{With SOC} & Sm-type & 64 & 70.06 & 4.598 & 4.598  & 18.121 & 90.00 & 90.00 & 90.69 \\
			 & Yb-mono & 12 & 0.000 & 4.689 & 4.573 & 17.411 & 90.00 & 87.38 & 90.00 \\
   
			\hline
            \multirow{2}{*}{Experimental}
			 & Bulk exp.\cite{nomura:1977} &  &  & 4.74 & 4.56 & 17.51 & & & \\
			 & This work &  &  & 4.73 & 4.57 & 17.50 & 90.0 & 86.85 & 90.0 \\
   
		\end{tabular}
	\end{ruledtabular}
\end{table*}  

Figure \ref{Fig3} summarizes the structural analysis of a 75~nm film grown via the optimized two-step procedure. Sharp, high-intensity Bragg peaks from the \{001\} \ce{LaBi2} lattice planes in the XRD $\theta/2\theta$ scan (\fig{3}a) indicate both considerable crystal quality and the strong X-ray scattering expected from heavy Bi ions. The spacing of the Bragg peaks corresponds to an out-of-plane lattice constant $c\approx 17.5$ \AA. Rocking curve analysis of the (006) diffraction condition (\fig{3}b) gives a FWHM $\approx 0.045^\circ$, indicating a uniform orientation of lattice planes throughout the film. $\varphi$-scans of both the \ce{LaBi2} film and MgO substrate in \fig{3}c demonstrate a [100]$_{\text{LaBi}_2}\parallel$ [110]$_{\text{MgO}}$ epitaxial register without 45$^\circ$-rotated domains.

Figure \ref{Fig3}d shows asymmetric reciprocal space mapping (RSM) taken along the $h=1$ rod (See Fig.~S4 of the Supplemental Material for extended RSM~\cite{Supplement}). As observed in epitaxial films of \ce{\textit{Ln}Sb2}, we observe a finite splitting of the ($\pm$1 0 12) peak along $Q_z$ associated with an out-of-plane tilt of the $\vec{a}^*$ reciprocal lattice vector. This implies a monoclinic structure for the \ce{LaBi2} thin films \cite{llanos:2024}, in contrast to the orthorhombic space group previously assumed in bulk crystal studies. The diffraction observed in RSM can be accurately indexed with in-plane lattice constants $a\approx 4.73$ \AA, $b\approx4.57$ \AA, and a monoclinic tilt of $\beta\approx86.9^\circ$ applied to one of three possible space groups (No.~12 \textit{A2/m}, No.~5 \textit{A2}, No.~8 \textit{Am}) \cite{tables}. While a definitive space group assignment based on XRD alone is not possible, supporting density functional theory (DFT) calculations of the thermodynamic stability of potential stacking configurations point towards the Yb-monoclinic structure as lower in energy than the competing Sm-type structure reported in bulk \ce{LaSb2} crystals \cite{llanos:2024}. The energy per La atom, the calculated lattice parameters both with and without SOC, and thin-film and bulk crystal experimental lattice parameters \cite{nomura:1977,petrovic:2002} are reported in Table \ref{tab:bulk}. The close agreement between these results suggests that the MBE-grown films stabilize in the same stacking configuration as bulk samples grown in Bi flux, which is in contrast to the case of \ce{LaSb2} \cite{llanos:2024}. The lattice parameters appear consistent with the Yb-monoclinic phase to within $\sim$~1$\%$. We therefore suspect that bulk studies have up to now misindexed monoclinic $\ce{LaBi2}$ as an orthorhombic system.  Due to the restrictions imposed by the samples' air sensitivity, we have been unsuccessful in directly imaging the crystal structure using electron microscopy. 

\begin{figure}[ht]
    \centering
    \includegraphics[width=85mm]{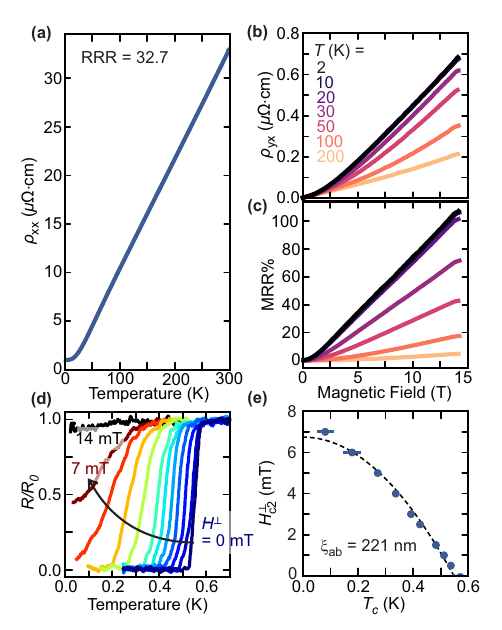}
    \caption{(a) Temperature-dependent sheet resistivity $\rho_\mathrm{xx}$ of a 75~nm film down to 2~K. (b) Hall resistivity $\rho_\mathrm{yx}$ and (c) symmetrized magnetoresistance ratio (MRR) under an out-of-plane field at a series of temperatures. (d) Superconducting transitions below 1 K plotted as the temperature-dependent ratio of resistance to normal-state resistance $R_{0}$, for a series of perpendicular magnetic fields. (e) Upper critical field $H_{c2}^{\perp}$ (defined at $R=0.5R_0$) versus temperature. Fit to the single-gap Ginzberg-Landau formula shown as dashed line.}
    \label{Fig4}
\end{figure}

\begin{figure}[ht]
    \centering
    \includegraphics[width=85mm]{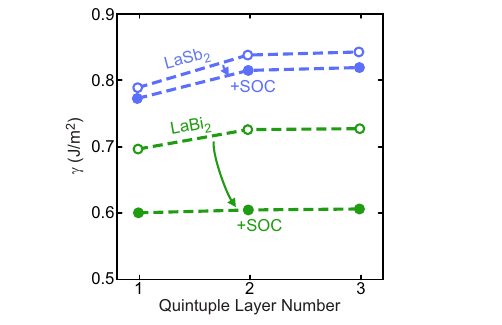}
    \caption{Calculated surface energy $\gamma$ as a function of the number of quintuple layers for \ce{LaSb2} (blue) and \ce{LaBi2} (green). The filled (open)
    circles represent the results obtained with (without) spin-orbit coupling (SOC).}
    \label{Fig5}
\end{figure}

A summary of magnetotransport measurements on the 75~nm thick \ce{LaBi2} film is presented in \fig{4}. The temperature-dependent sheet resistivity (\fig{4}a) shows metallic behavior with a RRR $(\rho_\mathrm{xx}(\text{300 K})/\rho_\mathrm{xx}(\text{2 K})$) of over 32. The anti-symmetrized Hall resistivity and symmetrized magneto-resistance ratio (MRR) are presented in \fig{4}b-c. The non-linear $\rho_\mathrm{yx}$ suggests at least two carrier types with a dominant hole-like species. The nearly linear, non-saturating MRR surpasses $100\%$ at 14~T.

Previous reports of \ce{LaBi2} transport properties were restricted to temperatures above the millikelvin regime. In \fig{4}d, we present resistance measurements down to $T$~=~20~mK where a sharp transition to a zero-resistance superconducting state is observed at about 550 mK at zero field. Suppression of this transition via a magnetic field applied perpendicular to the film confirms it as a transition into a superconducting ground state. Our phase-pure XRD analysis and UHV growth environment with persistent Bi desorption make it unlikely that this superconducting transition is a result of an extrinsic impurity phase; furthermore, reports of extrinsic superconductivity due to filamentary or thin-film Bi impurities in Bi-based compounds tend to occur at much higher temperature scales on the order of 1--2 K~\cite{xiang:2019}. We therefore conclude that this represents the first observation of intrinsic superconductivity in \ce{LaBi2} at ambient pressure. The superconducting coherence length is obtained by fitting the temperature-dependent critical field $H_{c2}(T)$ to the single-gap Ginzburg-Landau model,
\begin{equation}
    H_{c2}^{\perp}(T) = \left(\frac{\Phi_0}{2\pi\xi_{ab}^2}\right)\left( 1-\left(\frac{T}{T_c}\right)^2 \right).
\end{equation}

We find $\xi_{ab}\approx$ 221~nm. Based upon an estimate of the mean free path of 300~nm using the multicarrier Drude framework, we can infer that the films are likely clean-limit superconductors at this thickness.

\begin{figure}[ht]
    \centering
    \includegraphics[width=85mm]{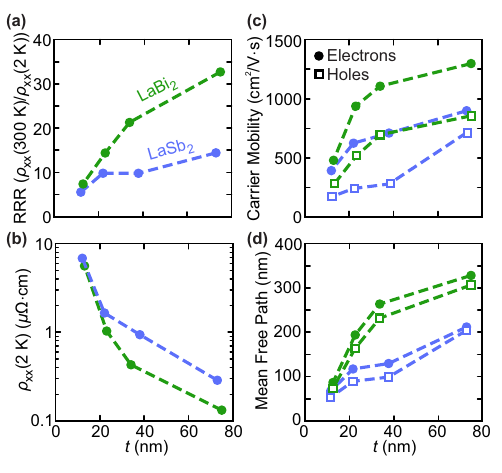}
    \caption{Comparison of experimental transport characteristics between \ce{LaSb2} and \ce{LaBi2} for several film thicknesses. (a) RRR and (b) base-temperature residual resistivity, the latter plotted on a log scale. Approximate (c) carrier mobility and (d) mean free path as estimated by fits of magnetotransport data to the two-carrier Drude model.}
    \label{Fig6}
\end{figure}

\section{Comparison of $\textbf{\ce{La}\textit{\ce{Pn2}}, \textit{\ce{Pn}} = \ce{Sb}, \ce{Bi}}$}

Despite the crystallographic similarities, there are notable differences between the growth dynamics and electronic properties of \ce{LaSb2} and \ce{LaBi2}, which will be the focus of the remainder of this manuscript. One notable difference is the strong RHEED oscillations observed during growth, as shown in \fig{2}, which are not observed in \ce{LaSb2} for any experimental condition. As RHEED oscillations are typically associated with a layer-by-layer growth mode, this contrast points to a stark difference in
the surface energies of the (001) lattice planes. To substantiate this analysis, density functional theory (DFT) calculations were performed to evaluate the surface energies of the LaSb- or LaBi-terminated corrugated layers as a function of quintuple layer (QL) thickness, where two QLs stacked along $\vec{c}$ comprise one conventional unit cell (See Fig.~S5 of the Supplemental Material~\cite{Supplement}). The results of these calculations are shown in \fig{5}. The surface energies are determined from~\cite{sun_efficient_2013,boettger_nonconvergence_1994} 
\begin{equation}\label{eqn:surface_energy}
	\gamma= {{E_\mathrm{s}-nE_\mathrm{bulk}}\over{2A}},
\end{equation}
where $E_\mathrm{s}$ is the total energy of the slab, $E_\mathrm{bulk}$ the bulk energy per atom, $n$ the number of atoms in the slab, and $A$ the surface area. The surface energies are found to converge rapidly at two QL, and two features stand out; \ce{LaBi2} has a smaller surface energy for all thicknesses than \ce{LaSb2},  and the inclusion of SOC significantly lowers these values, especially for \ce{LaBi2}.

\begin{figure*}[ht]
    \centering
    \includegraphics[width=170mm]{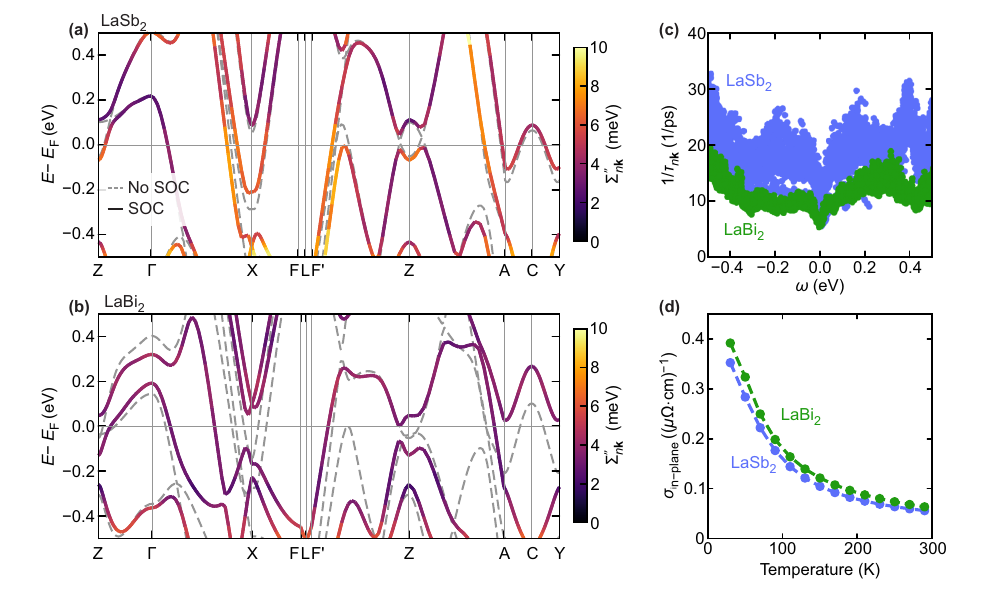}
    \caption{(a) $\mathrm{LaSb_2}$ and (b) $\mathrm{LaBi_2}$ band structures with (solid) and without (dashed) SOC. In the case of SOC, colors denote the band- and k-dependent $\Sigma_{n\textbf{k}}''$ which is proportional to the scattering rate. 
    (c) Calculated scattering rate for different band indices $n$ and $\mathrm{\textbf{k}}$ points, and 
    (d) temperature-dependent in-plane conductivity, $\sigma_\textrm{in-plane}=(\sigma_{xx}+\sigma_{yy})/2$ for \ce{LaBi2} (green) and \ce{LaSb2} (blue), respectively.}
    \label{Fig7}
\end{figure*}

The transport properties of \ce{LaSb2} and \ce{LaBi2} are compared in \fig{6} across several film thicknesses. The RRR (\fig{6}a) and low-$T$ residual resistivity (\fig{6}b) values are directly obtained from electrical measurements, whereas the  carrier mobilities (\fig{6}c) and mean free paths (\fig{6}d) are extracted from multicarrier Drude fits as described in the Fig.~S5 of the Supplemental Material~\cite{Supplement}. \ce{LaBi2} is observed to be more conducting than \ce{LaSb2} regardless of the metric used. While this may be intuitively attributed to an increased hopping integral for the enlarged Bi orbitals, our proceeding analysis of the band structure and phonon scattering rates provides an alternative insight into this behavior.

In order to understand the underlying origin of the transport properties of the two compounds, we have carried out {\it ab initio} calculations of the electrical conductivity tensor elements, $\sigma_{\alpha\beta}$, for the bulk phase, employing the self-energy relaxation time approximation (SERTA)~\cite{ziman_electrons_2001,ponce_epw_2016,giustino_electron-phonon_2017,ponce_towards_2018},
\begin{equation}\label{eqn:conductivity}
\sigma_{\alpha\beta} =  en_e \mu_{e,\alpha\beta} = -\frac{e^2}{\Omega} \sum_{n \in  \mathrm{CB}} \int_{\Omega_\mathrm{BZ}\in\mathrm{CB}} d\mathbf{k} \frac{\partial f_{n\mathbf{k}}^0}{\partial \varepsilon_{n\mathbf{k}}} v_{n\mathbf{k},\alpha} v_{n\mathbf{k},\beta} \tau_{n\mathbf{k}}^0,
\end{equation}

\noindent where $\alpha$, $\beta$ denote Cartesian coordinates, $n_e$ the electron carrier concentration, $e$ the electron charge, $\mu_{e,\alpha\beta}$ the electron mobility tensor component, $\Omega$ and $\Omega_\mathrm{BZ}$ the volumes of the unit cell and the first Brillouin zone, respectively, $\varepsilon_{n\mathbf{k}}$ the single-particle electron eigenvalue of band index $n$ and wave vector $\mathbf{k}$, $f_{n\mathbf{k}}^0$ the equilibrium Fermi-Dirac distribution function, $v_{n\mathbf{k},\alpha} = \hbar^{-1} \partial \varepsilon_{n\mathbf{k}}/\partial {k_{\alpha}}$ the $\alpha$ component of the group velocity, the sum runs over band indices $n$ within the conduction band (CB), and $\tau_{n\mathbf{k}}^0$ is the band- and $\mathbf{k}$-dependent electron-phonon (el-ph) relaxation time.

The scattering rate, 1/$\tau_{n\mathbf{k}}^{0}$, defined as the inverse of the relaxation time can be directly calculated from the imaginary part of the el-ph self-energy~\cite{ponce_towards_2018,ponce_epw_2016,giustino_electron-phonon_2017}, 

{\small
\begin{equation} \label{eqn:scattering_rate}
\begin{split}
1/\tau_{n\mathbf{k}}^0(\omega, T) = 2\Sigma''_{n\mathbf{k}}(\omega, T) 
= \frac{2\pi}{\hbar} \sum_{m\nu} \int_{BZ} \frac{d\mathbf{q}}{\Omega_\mathrm{BZ}} \left|g_{mn,\nu}(\mathbf{k}, \mathbf{q})\right|^2 \\
\quad \times \Big\{ \left[n_{\mathbf{q}\nu}(T) + f_{m\mathbf{k}+\mathbf{q}}(T)\right]\delta\left(\omega - (\varepsilon_{m\mathbf{k}+\mathbf{q}} - \varepsilon_F) + \omega_{\mathbf{q}\nu}\right) \\
\quad + \left[n_{\mathbf{q}\nu}(T) + 1 - f_{m\mathbf{k}+\mathbf{q}}(T)\right]\delta\left(\omega - (\varepsilon_{m\mathbf{k}+\mathbf{q}} - \varepsilon_F) - \omega_{\mathbf{q}\nu}\right) \Big\}.
\end{split}
\end{equation}
}

\noindent where $\omega$ is the energy of the electron, $T$ the temperature, $\omega_{\mathbf{q}\nu}$ the frequency, $n_{\mathbf{q}\nu}(T)$ the occupation (using Bose-Einstein statistics) of the phonon mode $\nu$ at wave vector $\mathbf{q}$, and $g_{mn,\nu}(\mathbf{k}, \mathbf{q})$ the electron-phonon matrix element between Kohn-Sham states ($n,\mathbf{k}$) and ($m,\mathbf{{k+q}}$) interacting with a phonon mode with wave vector $\mathbf{q}$ and band index $\nu$. 
The temperature dependence of the scattering rate arises from the electron and phonon occupations at finite temperatures, while electron-phonon matrix is calculated at $T$ = 0.

Figures \ref{Fig7}a and \ref{Fig7}b show the band structures of \ce{LaSb2} and \ce{LaBi2}, respectively, calculated with and without SOC. In both panels, the band structures are color-coded according to the imaginary part of the self-energy (i.e., inverse of the relaxation time). While \ce{LaSb2} exhibits relatively large scattering rates in some parts of the bands, \ce{LaBi2} shows overall smaller scattering rates throughout the Brillouin zone. 

To quantitatively corroborate the different scattering rates (and hence electron-phonon interactions) between the two systems, the scattering rate, $1/\tau_{n\mathbf{k}}^0(\omega, T=0)$, as a function of the chemical potential is collected over all band indices and sampled $\mathbf{k}$-points and displayed as a scatter plot, as shown in \fig{7}c.
This clearly demonstrates (i) substantially larger scattering rates in \ce{LaSb2} compared to those in \ce{LaBi2} over a wide range of chemical potential (and hence charge doping) and (ii) a wider distribution of scattering rates in \ce{LaSb2}, indicating the sensitivity of the relaxation time to both band index and the electron momentum. We attribute this difference in scattering rates between \ce{LaSb2} and \ce{LaBi2} to significant relativistic effects in \ce{LaBi2}. 

The calculated temperature-dependent in-plane conductivity, $\sigma_\textrm{in-plane} = (\sigma_\textrm{xx}+\sigma_\textrm{yy})/2$, is shown in \fig{7}d for \ce{LaSb2} and \ce{LaBi2}. As these calculations neglect the impact of static disorder, they cannot be compared directly to our experimentally obtained values. Nevertheless, they show that the relatively larger conductivity of \ce{LaBi2} over the entire temperature range considered arises from the lower scattering rates in \ce{LaBi2} compared to \ce{LaSb2}, rather than from differences in their relative group velocities (See Fig.~S8 of the Supplemental Material~\cite{Supplement}).
In summary, we attribute both key differences between the isostructural compounds \ce{LaBi2} and \ce{LaSb2} to the enhanced SOC in \ce{LaBi2} and its impact on the electronic structure. Our calculations show that the relativistic correction fundamentally alters the Fermi surface by shifting bands relative to the Fermi energy at high-symmetry points in the Brillouin zone. Consequently, the resulting energy landscape exhibits weaker phonon scattering and lower surface energy, which lead to more metallic transport and enhanced layer-by-layer growth in \ce{LaBi2} compared to \ce{LaSb2}.

\section{Conclusion}
We have reported the thin-film synthesis and the structural and electrical characterization of \ce{LaBi2}. Our structural analysis agrees well with bulk-reported lattice parameters but reveals a monoclinic structure, similar to thin films of \ce{LaSb2}~\cite{llanos:2024}, suggesting that previous bulk studies misindexed the compound. We have also demonstrated ambient-pressure superconductivity with $T_\mathrm{c} \sim 0.55$~K. Beyond these findings, this work elucidates the subtle role of SOC in influencing both thin-film growth dynamics and metallicity. Compared to \ce{LaSb2}, SOC in \ce{LaBi2} lowers the (001) surface energy and suppresses carrier scattering near the Fermi energy, resulting in layer-by-layer growth and superior electrical conductivity.

\section{Acknowledgments}
This material is based upon work supported by the National Science Foundation Graduate Research Fellowship under Grant No. 2139433. We acknowledge funding provided by the Gordon and Betty Moore Foundation’s EPiQS Initiative (Grant number GBMF10638), the Institute for Quantum Information and Matter, a NSF Physics Frontiers Center (NSF Grant PHY-2317110). The work at CSUN was supported by the NSF-PREP CSUN/Caltech-IQIM Partnership (Grant No. 2216774).  The calculations were carried out in an Advanced Beowulf Cluster funded by the NSF grant number DMR-2406524. S.-H.J. was supported by Basic Science Research Institute Fund (NRF grant number RS-2021-NR060139).

\clearpage 

\bibliography{bib}
\clearpage
\title{Relativistic Effects in \ce{LaBi2} Thin Films\\
Supplementary Material} 
\author{Reiley~Dorrian}
\affiliation{Department of Applied Physics and Materials Science, California Institute of Technology, Pasadena, California 91125, USA.}
\affiliation{Institute for Quantum Information and Matter, California Institute of Technology, Pasadena, California 91125, USA.}

\author{Sungmin~Song}
\affiliation{Department of Physics, Pohang University of Science and Technology, Pohang 37673, Republic of Korea}
\affiliation{Department of Physics and W. M. Keck Computational Materials Theory Center, California State University Northridge, Northridge,
California 91330, USA.}

\author{Jinwoong~Kim}
\affiliation{Department of Physics and W. M. Keck Computational Materials Theory Center, California State University Northridge, Northridge,
California 91330, USA.}
\affiliation{Korea Institute for Advanced Study, Seoul 02455, Republic of Korea}

\author{Mizuki~Ohno}
\affiliation{Department of Applied Physics and Materials Science, California Institute of Technology, Pasadena, California 91125, USA.}
\affiliation{Institute for Quantum Information and Matter, California Institute of Technology, Pasadena, California 91125, USA.}


\author{Seung-Hoon Jhi}
\affiliation{Department of Physics, Pohang University of Science and Technology, Pohang 37673, Republic of Korea}

\author{Nicholas~Kioussis}
\affiliation{Department of Physics and W. M. Keck Computational Materials Theory Center, California State University Northridge, Northridge,
California 91330, USA.}

\author{Joseph~Falson}
\email{falson@caltech.edu}
\affiliation{Department of Applied Physics and Materials Science, California Institute of Technology, Pasadena, California 91125, USA.}
\affiliation{Institute for Quantum Information and Matter, California Institute of Technology, Pasadena, California 91125, USA.}

\maketitle

\renewcommand{\thefigure}{S\arabic{figure}}

\setcounter{figure}{0}

\textit{Multi-carrier fitting}:
The estimated carrier mobilities reported in the main text were obtained from fits of the magnetoresistance and Hall resistivity (\fig{S6}) to the two-carrier Drude formulas,
\begin{align}
    \rho_{xx} &= \frac{1}{e}\frac{(p\mu_h+n\mu_e)+(p\mu_e+n\mu_h)\mu_h\mu_eB^2}{(p\mu_h+n\mu_e)^2+(p-e)^2\mu_h^2\mu_e^2B^2}\\
    \rho_{yx} &= \frac{B}{e}\frac{(p\mu_h^2-n\mu_e^2)+(p-n)^2\mu_h^2\mu_e^2B^2}{(p\mu_h+n\mu_e)^2+(p-e)^2\mu_h^2\mu_e^2B^2}
\end{align}
where $n$ ($p$) and $\mu_e$ ($\mu_h$) are the carrier density and mobility of electron-like (hole-like) carriers, respectively, $B$ is the applied magnetic field, and $e$ is the electron charge. Fits are performed to both $\rho_{xx}(B)$ and $\rho_{yx}(B)$ simultaneously, with the additional constraint that the zero-field conductivity agrees with $\sigma_{xx}(B=0) = e(n\mu_e+p\mu_h)$.

\textit{Orbital-projected band structures}:
Figure \ref{FigS7} presents the orbital-projected band structures of \ce{LaBi2} with and without spin-orbit coupling (SOC). 
Significant SOC due to the heavy mass of bismuth induces hybridization of the La-$d$ and Bi-$p$ orbitals, modifying the overall shape of the band structure. 
Consequently, electron and hole pockets disappear in the vicinity of the zone boundaries (specifically, near $\mathrm{X}$ and $\mathrm{C}$ points), thus significantly altering the intrinsic electron-phonon scattering conditions.

\textit{Electronic group velocities}:
The magnitudes of the group velocities, $|v_\mathrm{g}|$, projected onto the band structures of \ce{LaSb2} and \ce{LaBi2} are shown in \fig{S8}. \ce{LaBi2} exhibits smaller group velocities than \ce{LaSb2} due to the band modification induced by SOC, specifically the gap opening near the $\mathrm{X}$ point and the lifting of electron/hole pockets along the $\mathrm{A-C-Y}$ path. This confirms that the higher conductivity of \ce{LaBi2} originates from the relaxation time $\tau^0_{n\mathbf{k}}$, which compensates for the relatively small group velocity $|v_g|$.

\newpage
\begin{figure}[h]
    \centering
    \includegraphics[width=160mm]{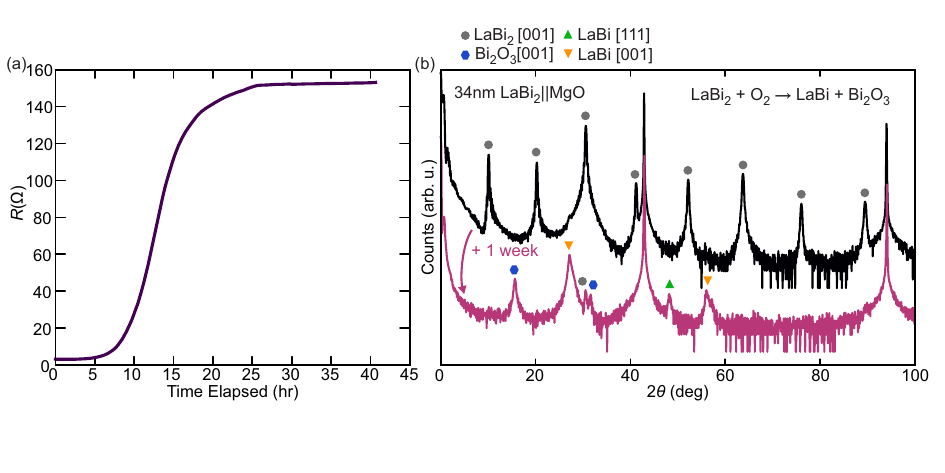}
    \caption{Degredation of a Ge-capped \ce{LaBi2} film through air exposure. (a) Four-point resistance of a $\sim0.5\times1$cm sample versus time in air. Resistance slowly rises during the first few hours of exposure, then degradation rapidly accelerates and saturates after roughly 24 hours. (b) $\theta/2\theta$ scans showing phase composition of a film before and after complete degradation. \ce{LaBi2} peaks are diminished in place of multiply-oriented LaBi and \ce{Bi2O3}.}
    \label{FigS1}
\end{figure}

\begin{figure}[h]
    \centering
    \includegraphics[width=80mm]{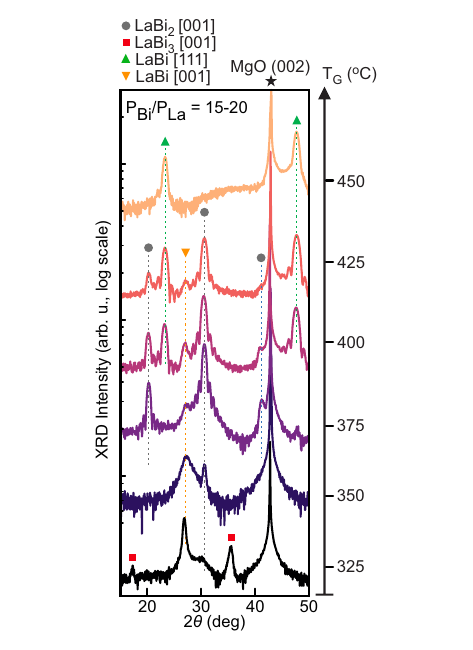}
    \caption{$\theta/2\theta$ diffraction patterns resulting from single-step growth at fixed substrate temperature $T_\text{G}$ for 1 hour. Maximum $T_\text{G}=450^\circ$C and minimum $T_\text{G}=325^\circ$C reported in the main text where excessive Bi desorption and limited migration energy at the substrate surface, respectively, result in dominant impurity phases. A small window around $375^\circ$C shows the most phase-purity of the series albeit with a noticeable LaBi impurity still present. Thus a two-step growth procedure is a necessity to produce single-phase \ce{LaBi2} thin films of appreciable thickness.}
    \label{FigS2}
\end{figure}
\newpage

\begin{figure}[h]
    \centering
    \includegraphics[width=170mm]{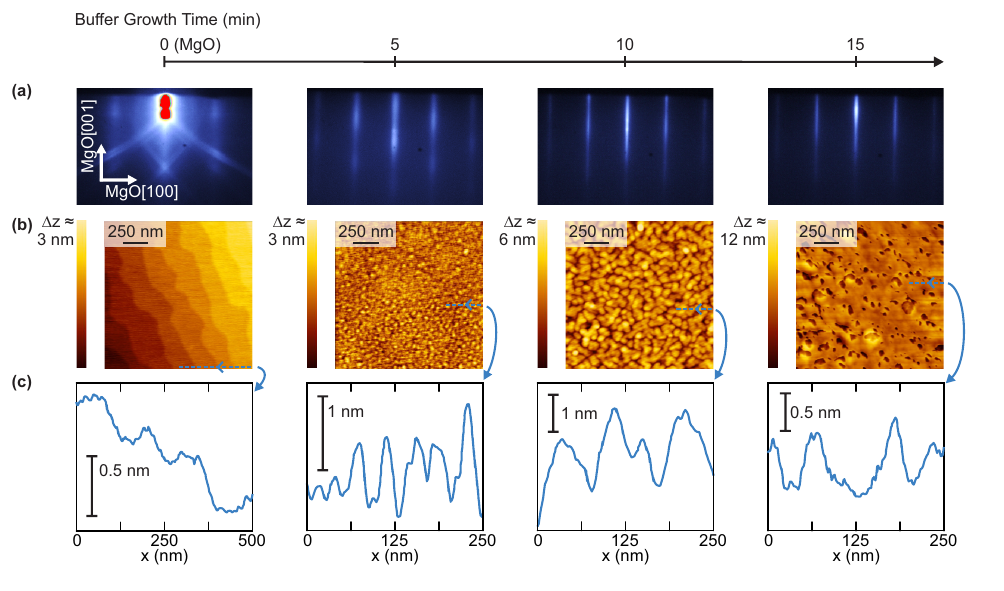}
    \caption{Progression of buffer layer growth at $T_\text{G}\approx 450$-$475^\circ$C viewed by (a) \textit{in-situ} RHEED and (b) atomic force microscopy (AFM, performed in an \ce{N2}-environment glovebox). To better represent the morphology of the samples, representative line-cuts are shown in (c). The step-terraced MgO substrate becomes decorated with $\sim$1 nm tall islands in the first 5 minutes of growth, which gradually merge as the film wets the substrate surface. 
    By 15 minutes of growth the film becomes flat and contiguous besides some residual pits where the film has yet to merge.}
    \label{FigS3}
\end{figure}

\newpage
\begin{figure}[h]
    \centering
    \includegraphics[width=170mm]{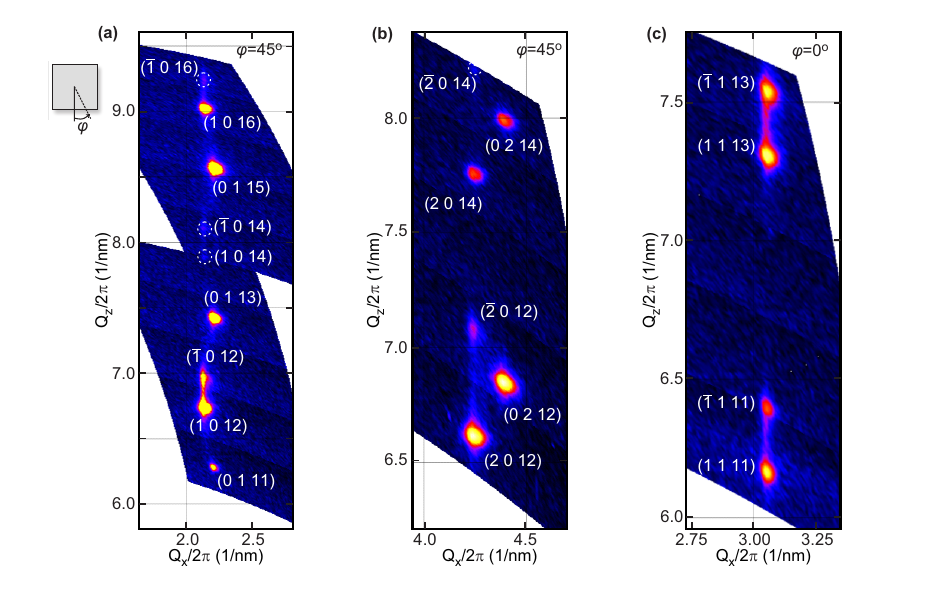}
    \caption{Extended reciprocal space mapping along the (a) $h=1$, (b) $h=2$, and (c) $h=k=1$ rods. Splitting along $Q_z$ is observed for all diffractions with non-zero $h$. These three scans along with the $\theta/2\theta$ pattern indexed in the main text allow full identification of the structure's extinction coefficients; $\ell = 2n$ for $(00\ell)$; $k+\ell = 2n$ for $(0k\ell)$; $\ell = 2n$ for $(h0\ell)$; $k+\ell=2n$ for $(hk\ell)$. This is consistent with three possible monoclinic space groups (SG 5, 8, and 12), between which space group 12 is identified via first-principles DFT calculations.}
    \label{FigS4}
\end{figure}

\newpage
\begin{figure}[h]
    \centering
    \includegraphics[width=170mm]{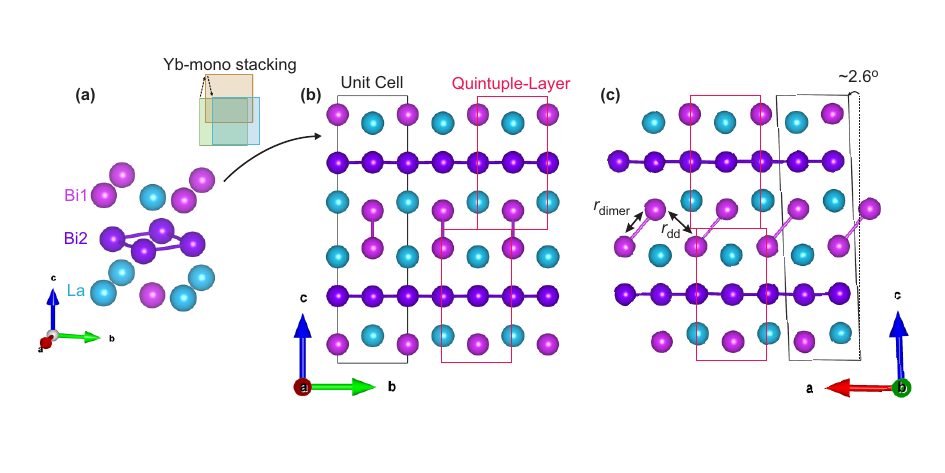}
    \caption{Yb-monoclinic structure of the \ce{LaBi2} in terms of the quintuple-layer construction. (a) A single quintuple-layer block contains one distinct La and two distinct Bi sites. The Yb-mono structure is obtained by a staggered stacking configuration of these blocks~\cite{llanos:2024}. The resulting structure is shown along the (b) $\vec{a}$ and (c) $\vec{b}$ crystallographic directions.}
    \label{FigS5}
\end{figure}

\begin{figure}[h]
    \centering
    \includegraphics[width=160mm]{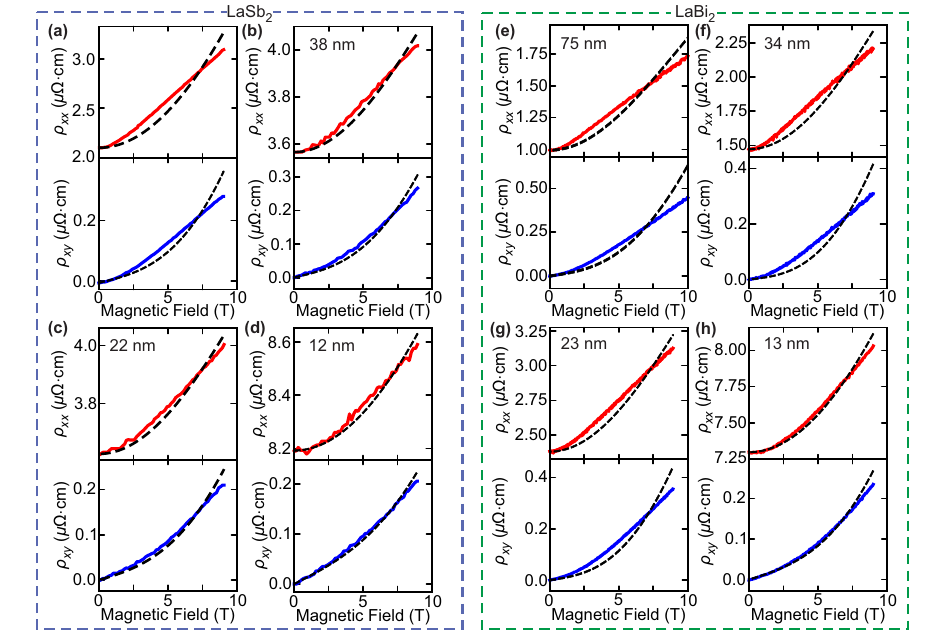}
    \caption{Fitting of the two-carrier Drude model to longitudinal magnetoresistance and hall resistivity data for (a-d) \ce{LaSb2} and (e-h) \ce{LaBi2}, for four representative film thicknesses. Significant deviations from the model are apparent in the thicker films primarily due to the non-quadratic behavior of the magnetoresistance. However, these deviations are apparent in both compounds so direct comparison of the fit parameters between both series is acceptable for the purpose of this study.
    }  
    \label{FigS6}
\end{figure}

\newpage
\begin{figure}[h]
    \centering
    \includegraphics[width=170mm]{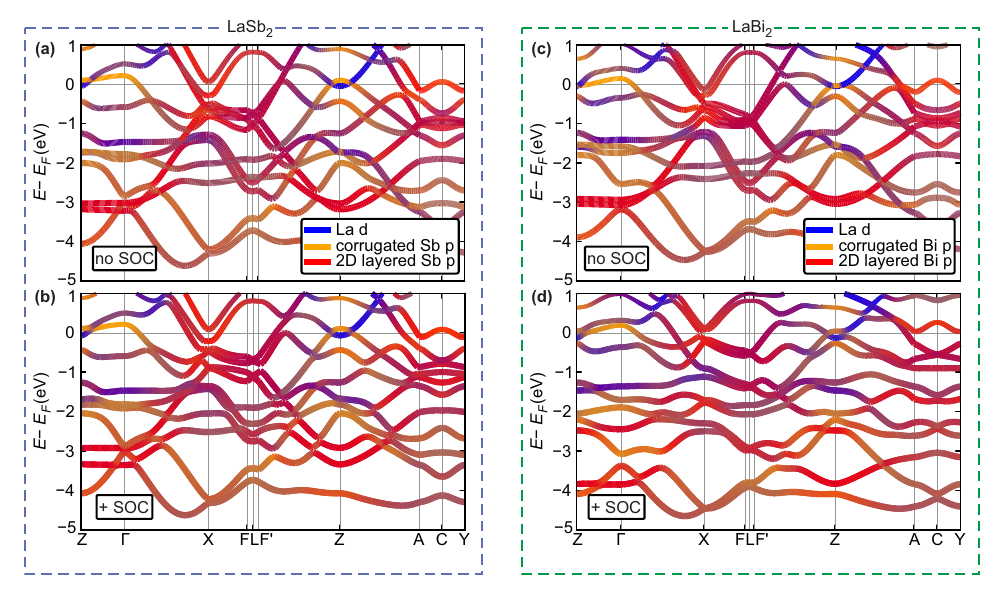}
    \caption{The orbital-projected band structures of (a, b) \ce{LaSb2} and (c, d) \ce{LaBi2} without and with SOC. }
    \label{FigS7}
\end{figure}

\newpage
\begin{figure}[h]
    \centering
    \includegraphics[width=85mm]{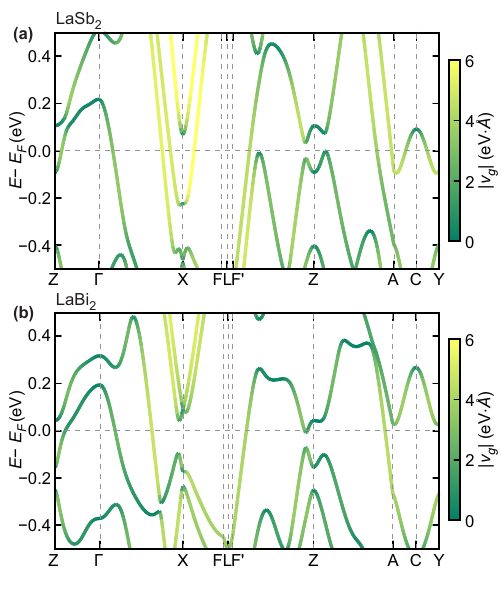}
    \caption{The band structures with SOC projected by electronic group velocities of (a) LaSb2 and (b) LaBi2. }
    \label{FigS8}
\end{figure}


\end{document}